# Optimizing broad ion beam polishing of zircaloy-4 for electron backscatter diffraction analysis

Ning Fang[1*], Ruth Birch[1*], and Ben Britton[1,2+]

1. Department of Materials, Imperial College London

2. Department of Materials Engineering, University of British Columbia

* Joint first author

+ Corresponding author: ben.britton@ubc.ca

**Lay Abstract**

To provide useful materials characterization, we must prepare samples well so that we can avoid studying artifacts induced during sample preparation and this motivates us to systematically study our preparation methods. In this work, we focus on improving "broad ion beam" (BIB) polishing through a combination of consideration of the ion-sample interactions and a systematic study of options provided by commonly available broad ion beam milling machines. Our study specifically aims to optimize the preparation of zircaloy-4, which is an alloy of zirconium used in nuclear fuel cladding, and we note that this alloy is difficult to prepare with other sample preparations routes. We optimize BIB polishing to study the microstructure of the zircaloy-4 with electron microscopy based electron backscatter diffraction (EBSD). To conclude our study, we provide recommendations for new users of BIB based polishing methods.

**Abstract**

Microstructural analysis with electron backscatter diffraction (EBSD) involves sectioning and polishing to create a flat and preparation-artifact free surface. The quality of EBSD analysis is often dependant on this step, and this motivates us to explore how broad ion beam (BIB) milling can be optimised for the preparation of zircaloy-4 with different grain sizes. We systematically explore the role of ion beam angle, ion beam voltage, polishing duration and polishing temperature and how this changes the surface roughness and indexing quality. Our results provide a method to routinely prepare high-quality Zircaloy-4 surfaces, and methods to optimise BIB polishing of other materials for high-quality EBSD studies.

I. **Introduction**

The quality of understanding of materials depends on how well we can characterise them, and this is influenced significantly by the quality and method of sample preparation. For metallographic examination, typically a high-quality flat mirror surface is required for analysis. To achieve this a sample is sectioned mechanically, mounted, and the section surface is mechanically ground and polished. For metals, mechanically sectioning, grinding and polishing introduces surface damage, including an amorphous layer and cold work (i.e. dislocations). Removing this damaged layer and revealing the pristine sample surface are essential for high-resolution characterisation with electron microscopy, as the electrons typically only penetrate the surface layer, which is typically less than 1 µm, and ~20 nm for electron backscatter diffraction, EBSD analysis.

To remove this surface layer without causing more damage, often smaller and smaller grit sizes are used, or a chemical or ion polishing step is used. For chemical polishing, electropolishing can be used but this often involves the use of hazardous chemicals and can be extremely difficult to perform routinely across different alloys, as the electropolishing chemistry and conditions depend on the chemistry of the particular alloy and the precise chemistry of the (ageing) chemical polishing solution.

Alternatively, gentle removal of the surface layer can also be performed using the ion milling process, where high-energy ions (>0.5 keV) are used to sputter and remove the surface layer. This approach typically causes less damage than mechanical methods [1], and avoids the production of hydrides induced by mechanical grinding and polishing in acidic conditions [2].

For large-area samples, broad ion beam (BIB) processing has been developed, where a broad beam of charged (usually $Ar^+$) ions are focussed at an inclined angle (0-18° from parallel) to the sample surface. For some materials, it can replace mechanical polishing methods and save hours of mechanical polishing time[3]. It has also been extended to support destructive 3D sectioning to enable large-area 3D analysis with high indexing rates [4].

In BIB, energetic ions are used to remove the surface, and they can also cause surface radiation damage. Our goal is to maximise the even removal of our sample surface, while reducing the ion beam induced radiation damage.

Ion beam damage is caused by direct implementation, as well as when the incident ions knock lattice atoms/ions from their parent sites. This process is called the 'collision' cascade.

For ion beam polishing, the incoming ion beam must have enough energy such that a large fraction of these ions can remove preparation induced damage. Each ion will interact with the lattice and it can: (1) cause sputtering - i.e. the liberations of parent material to vacuum, typically at the surface layer; (2) scatter elastically or near elastically, causing no damage but potentially slowing the ion down, reducing its energy, and changing its path; (3) knock a parent lattice ion from its site and create a Frenkel pair (a self-interstitial and a vacancy); (4) run out of energy and stop in the lattice [5,6].

When the parent ion is knocked from its site, the Frenkel pair will often, at temperatures above 0K, anneal and recover. In some cases, these point defects can remain (i.e. increase in point defect density in the surface region); they may cluster to form prismatic dislocations; and they may support diffusion or changes in the chemistry of the parent material (e.g. supporting the accumulation of hydrogen ions). The likelihood of each of these events is controlled by material, incoming ion species and energy and dose rate. The atomic diffusion and transport mechanisms are strongly temperature dependant.

For BIB polishing, the quality of the surface is typically controlled by beam energy, ion current and angle of incidence [7] and these parameters motivate our systematic exploration in the present work.

In most BIB equipment, the gun current can be measured periodically using Faraday cups located opposite the guns in the work chamber or it can be determined from the accelerating current. Measurement of the current is useful as the total ion dose on the sample is a product of the polishing time, gun focus, and gun current. The dose rate is controlled by the current, as well as how the sample is moved under the broad ion beam.

The beam energy is controlled by varying the accelerating voltage. The voltage provides more energy to the large ions, and this can encourage a higher amount of sputtering. However, a higher energy ion can implant further and create more knock-on damage, which may result in the generation of radiation damage (e.g. phase transformations, point defects or clustering of point defects into prismatic loops).

In addition to sputtering of the base material, the ion beam can also clean surface hydrocarbons and remove any surface layers (e.g. oxides) which can improve the surface quality significantly.

Polishing using broad ion beam milling is typically performed when the sample surface is nearly parallel (0-18 °) to the incident ion beam. This reduces direct implantation significantly, but it can also result in the shadowing and curtaining ('waterfall' effects). These are created when the incoming beam is blocked or the milling is incomplete, e.g. due to prior surface roughness or heterogeneity in the milling and sputtering rate of the surface [8]. Furthermore, for broad ion beam polishing, Datta et al. [9] show the evolution of the periodic ripple patterns, whose orientation are perpendicular to the projection of the ion beam within the incident angle ranging from 55° to 70° with respect to the sample normal (i.e. 10 to 25° from parallel to the surface plane, i.e. any vector in the surface of the sample). Datta et al. also show that this roughness can often be made worse when the polishing time is increased, prior to reaching a steady state. These surface roughness artifacts can be reduced if the sample is moved under the beam, e.g. using sample rotation to avoid the growth of (self) shadow-induced and curtaining [1,10].

The temperature of the sample during ion beam polishing can be used to control the types and extent of radiation damage formed. The mechanisms here are less clear, but they are related to the combination of surface sputtering and the creation of point defects, diffusion of chemical species and re-oxidation of the surface. These chemical degradations and changes in microstructure typically increase within increasing temperature [11–15].

For example, cryo-preparation of atom probe tomography needles using (focussed) ion beam methods has been shown to reduce contamination with hydrogen [16,17]. At room temperature, the ion beam milling leads to the formation of undesired hydrides induced by hydrogen pick-up during the preparation of TEM lamellae in Zr alloys [17,18]. Furthermore, Chang et al. and Rivas et al. demonstrate that the cryo-milling can prevent the formation of defects and preserve the pristine nature of sensitive materials [16,17]. Desbois et al. also combine BIB with cryogenic temperature (cryo) facilities to produce the large, high-resolution cross-sections in SEM at a cold temperature [19]. This novel BIB-cryo-SEM enabled the production of 2D cross-sections without noticeable curtaining [20].

While BIB has significant potential for the preparation of samples for metallographic examination, it remains challenging to employ a wide range of materials as the polishing recipes are less well explored and so the design of the polishing step needs to be considered. The BIB machine needs to be designed to minimise contamination, reduce gun cleaning (which affects the ion current and probe shape) and optimisation of these recipes requires knowledge of ion-sputtering and damage behaviour [7].

Many of the advances in BIB used for TEM lamella preparation can be extended to support the development of SEM sample preparation recipes.

Furthermore, the wide adoption of focussed ion beam (FIB) instruments has also resulted in advances in the understanding of ion-beam induced damage. FIB preparation routes can be good where BIB instruments are not available but FIB surface polishing is typically limited to small areas of 100–200 $\mu m^2$ [1], as compared to 1 $mm^2$ for BIB. Furthermore, the $Ar^+$ ions used in BIB are larger than $Ga^+$ ions, and so can effectively sputter significant volumes of material with minimal ion damage (e.g. point defect generation) and limited changes in surface chemistry, as compared to $Ga^+$ beams [1,21–23]. The depth of the FIB-induced damaged layer is about 10 nm, which is approximately 3 times larger than the depth of the BIB-induced damaged layer [10].

However, we note that we can learn about successful BIB polishing routes based upon extensive experiences from the Ga-FIB community. Furthermore, we can use these techniques in unison, such as cleaning samples with BIB after FIB fabrication [23] to reduce the thickness of the FIB-induced damaged layer [21].

It remains challenging to use BIB to provide site-specific samples, as the area polished is large. Cross-sectional samples can be made, e.g. by positioning a radiation hard blade on the surface of the sample and milling down from there. Other advanced uses of BIB milling include 3D-destructive investigations where characterisation is performed after each BIB layer reduction. This is described by (Dieterle, Butz, and Müller 2011; Winiarski et al. 2017) for 3D-EBSD to explore large areas (>1mm$^2$) similar to the work of Hosman et al.[24] to provide 2D EBSD maps for large areas, and 3D tomographs can be constructed from repeat sectioning [4,24].

BIB has been used widely for sample preparation, but there are limited systematic studies of its use and development of the best polishing conditions. In this work, we focus on using BIB polishing to prepare Zircaloy-4 microstructures which are challenging to routinely prepare using conventional routes. We focus on exploring the role of the beam voltage, gun angle, polishing duration and specimen temperature to optimise EBSD analysis.

## II. Experimental procedure/ Methods

Samples of fine grain zircaloy-4 (Zr4, Zr-1.5Sn-0.2Fe-0.1Cr, wt%) were sectioned from a large plate of zircaloy-4. These samples were metallographically prepared up to a 2000 grit SiC finish. Samples left after this stage are the 'ground samples'.

Large grain samples were prepared by taking some of the fine grain samples and applying a 'blocky alpha' heat treatment which involves holding them for 336 hours (14 days) at 800 °C in argon atmosphere based upon the recipe developed by Tong [25,26].

These large grain samples were ground to a 2000 grit SiC finish, then were additionally polished using a silica slurry which was mixed with hydrogen peroxide and water to balance the pH and control the colloid behaviour (5 OPS: 1 hydrogen peroxide: 6 water, by volume).

Key microstructural features for both sample types are shown in Figure 1. Note that both samples contain a dispersion of fine secondary phase particles which are well distributed within the much large α-zirconium grains. Analysis of the EBSD maps from the fine grain samples have an average (arithmetic) grain size of 12.1±3.57 µm and that these large grain samples typically have an average grain size in excess of 100 μm, with some smaller 'island' grains (often ~20 μm).

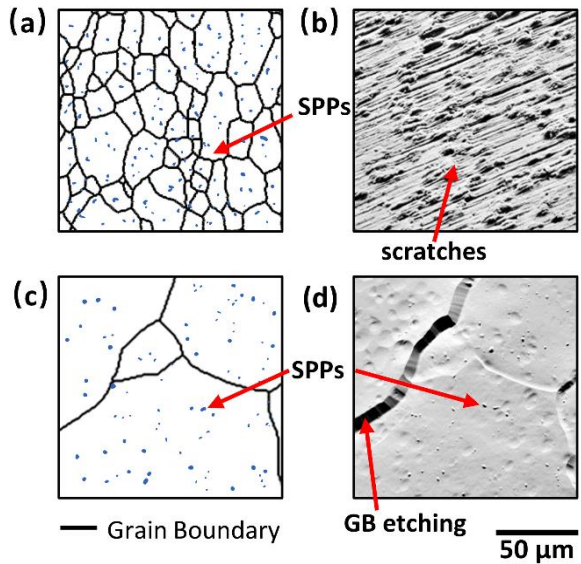

*Figure 1 Microstructural features in the Zircaloy-4 samples. (a) schematic of the fine grain sample; (b) optical micrograph of as ground fine grain sample; (c) schematic of the large grain microstructure; (d) scanning electron micrograph of the large grain sample after broad ion beam polishing. [The magnification is the same for all the micrographs in this figure.]*

Broad ion beam polishing experiments were performed using a Gatan Model 685 Precision Etching and Coating System (PECS II) (Gatan, Inc. Pleasanton, USA), as shown in **Error! Reference source not found.**(a). This instrument has two Ar-ion guns that are aimed at a sample which can rotate while the system is under vacuum, and the system can be operated at low temperature using a cold finger chilled by liquid nitrogen, connected via conductive brushes. The PECS II ion guns can vary in voltage from 0.5 keV to 8 keV.

The user can easily vary the angle of the two ion guns independently between 0 and 18°, where 0° has the ion gun pointing along the sample surface, as shown in **Error! Reference source not found.** (b). During polishing, the sample is rotated about the coincident point to reduce ion beam shadowing.

The two guns are typically focussed to be co-incident and the width of each beam is typically 3 mm. Due to challenges in the operation of the instrument, the two beams may be misaligned, not perfectly focussed and have a different ion beam current. This can affect the dose, especially as the beams are elongated when they are incident on the tilted sample.

To perform ion polishing at a low temperature, the sample is chilled using liquid nitrogen (LN2). In the PECS II system, users fill the LN2 dewar prior to loading a sample into the chamber. Samples can be added to the chamber after 30 minutes and stay in the chamber for more than 30 minutes. The sample takes time to cool and waiting longer between adding the liquid nitrogen and polishing results in a colder sample. The manual for the PECS indicates that a ~near steady state is reached after 120 minutes, and this is typically -160 °C. If samples are cooled, a cold sample should be left in the chamber after polishing and the stage heater will be turned on for warming up before removing the samples to prevent condensation.

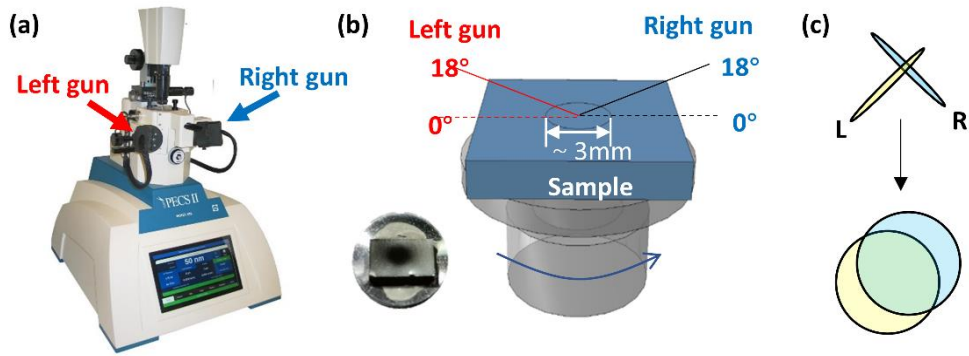

*Figure 2 (a) Front view of PECS II system (Gatan Inc., 2014); (b) Area of the sample surface polished and the range of the gun tilting angle; (c) Cartoon showing how gun misalignment can vary the polished area.*

For the present work, sample preparation with the broad ion beam system was conducted using different sample preparation recipes and the parameters explored (and other conditions) are shown in Table 1. These include varying the angle of the guns, the accelerating voltage, the polishing duration, and the temperature of the sample.

| Trial Name | Number of Samples | | Conditions (with 1 rpm and no modulation) | | | |
|---|---|---|---|---|---|---|
| | Fine Grain | Large Grain | | | | |
| **Angle** | 5 | 5 | 8 keV | 2, 4, 6, 8,12 ° | RT | 15 mins |
| **Voltage** | 3 | 3 | 2, 4, 6, 8 keV | 6 ° | RT | 15 mins |
| **Duration** | 1 | 0 | 8 keV | 8 ° | RT | 3.75, 7.5, 15, 30, 60 mins |
| **Cold** | 1 | 1 | 8 keV | 8 ° | RT & Cold | 15 mins |
| **Cold + Voltage** | 0 | 4 | 2, 4, 6, 8 keV | 6 ° | RT & Cold | 15 mins |

*Table 1 Outline of experiments conducted number and recipes of fine and large grain samples.*

EBSD analysis was performed on a Quanta 650 FEG SEM (Thermo Fisher Scientific, Waltham, USA) equipped with a Bruker HR detector. Maps were collected at 20 kV, a working distance of 18 mm, and a typical probe current of 10 nA. The EBSD detector was inserted to a detector distance of 17 mm. For each sample, representative EBSD maps were collected with a step size of 0.86 µm with 500 x 333 points per map and patterns were collected with 160 x 120 pixel resolution EBSPs with an exposure time of 10ms per pattern. Patterns were indexed online in eSprit v2.3, and they were indexed using the as supplied α-Zr HCP phase. Dynamic background correction was used, and patterns were collected in a 16-bit format.

The Bruker eSprit software includes two quality metrics: (1) the 'hit rate' (HR) which is defined as the number of indexed points divided by the number of possible points in the selected area, where each point is indexed if the required minimum number of bands have been indexed (five in this case); (2) the 'Mean Radon Quality' which is the height of the identified Hough peaks, with up to 8 peaks used and the standard Radon settings employed (60 Hough Resolution), divided by the average intensity in the Radon background and then subsequently normalised to a scale of 0-1.

The EBSD camera has three fore scatter diodes (FSDs) attached to form a false colour RGB microstructural image. When the camera is inserted for good EBSD pattern collection, these FSD images tend to highlight surface topography [27].

Surface topography was measured with white light interferometry using a NewView 7000 (Zygo Corporation, Middlefield, USA). Images were analysed to explore surface roughness using the arithmetic mean of the surface height (Sa), the root mean square height (Sq) and the range of heights (Sz), as measured from the difference of the average of the 5 highest points and the 5 lowest points [28]. As we are typically exploring the variation in local surface roughness, we focus on the variation of Sq with polishing conditions.

## III. Results

The effects of changing the angle, voltage, polishing duration and temperature are explored using analysis of data collected with SEM-based imaging, EBSD mapping, and surface roughness. In this results section, each experiment trial (detailed in Table 1) is explored in turn.

### 1. Angle trials

The gun tilts were varied from 2 to 12° (other conditions are shown in Table 1) and representative FSD and EBSD maps are shown in Figure 3, with a numerical assessment of the hit rate, radon quality, and the surface roughness is presented in Figure 4. For both grain sizes, a higher gun tilt angle results in a higher indexing success, with an angle of >8° sufficient to achieve >99% indexing success for both samples, and the large grain sample indexed successfully with an angle >4°.

The FSD images accentuate imaging of surface roughness, and when compared with the EBSD maps they show that there is roughness associated with specific features, such as grain boundaries and the secondary phase particles, and this is especially visible within the large grain sample.

In all cases, the EBSD pattern quality, on average, increases within the increased polishing angle. However, in both cases, there are systematic areas where there is poor indexing.

In the fine grain samples, grains with their <c> axis pointing out of the plane of the sample tend to polish faster than other grains (i.e., the index IPF-Z maps are 'redder' than expected). As the quality of the polishing improves, regions of poor indexing become associated only with small regions spaced at high spatial frequency, which is probably associated with different polishing rates and near secondary phase particles.

In the large grain sample, poor indexing (and variations in roughness) is associated with the grain boundary region and there are specific regions where poor quality EBSD patterns are located up to 5 µm from the grain boundary, as seen in the 4° polished large grain sample. In the higher angle maps, the grain interior is smooth while the grain boundaries are highly stepped. For the very high angle polish, high-frequency regions of poor indexing are seen, and these are probably correlated with preferential etching associated with secondary phase particles.

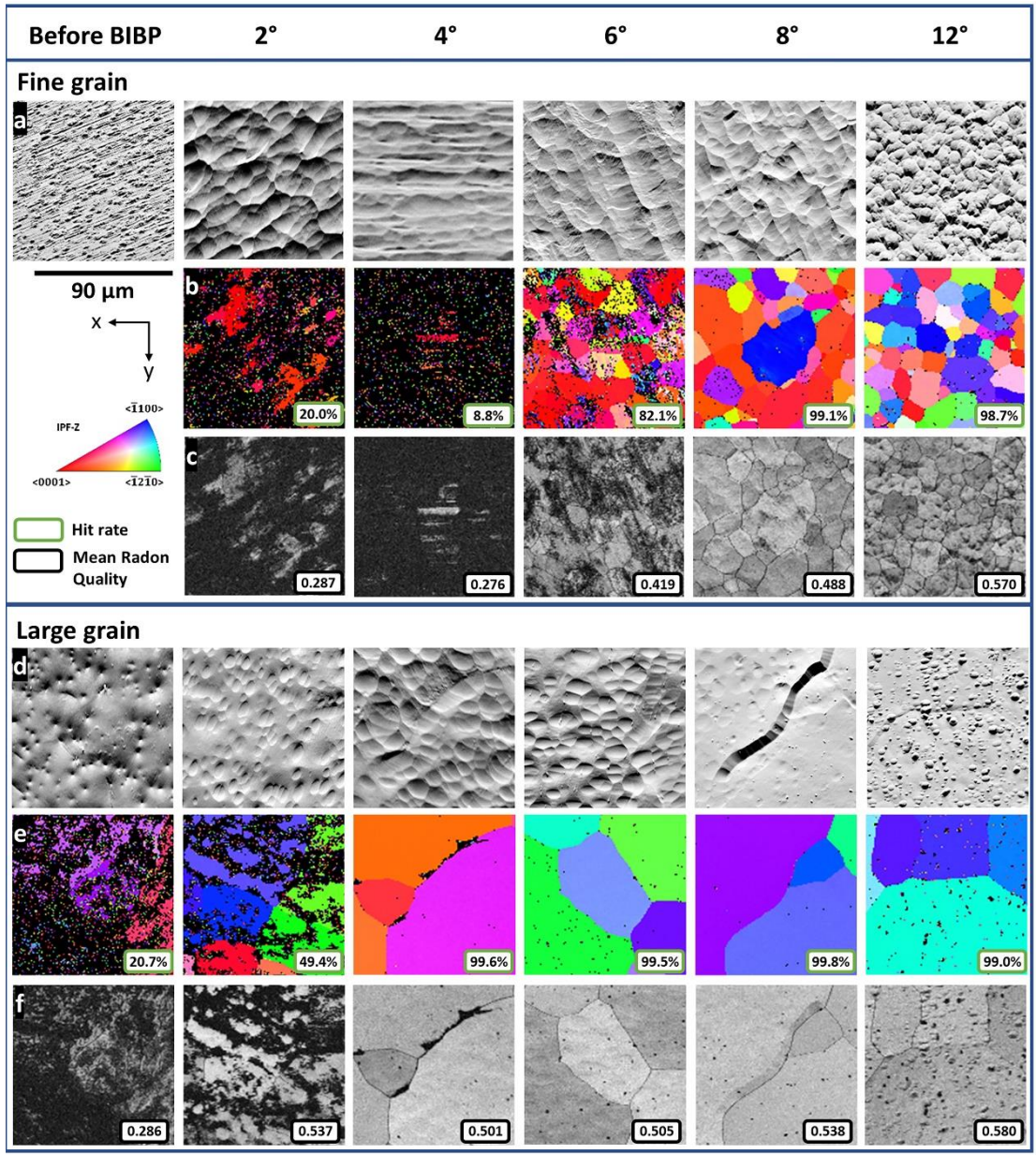

Figure 3 Effect of angle on broad ion beam polishing quality for successful EBSD mapping and forescatter imaging of large grain (OPS polished) and fine grain (ground to 2000 grit using SiC) Zircaloy-4. The numbers in the green box are the percentage of successfully indexed points, and the numbers in the black box are the mean radon quality. Recipe: 8keV, no modulation, 1 rpm, 15 mins, 2-12 ° beam angle, room temperature. (a)(d) Forescatter imaging, (b)(e) successful EBSD mapping with IPF-Z colouring and non-indexed points in black, (c)(f) normalised pattern quality.

[The magnification is the same for all the micrographs in this figure.]

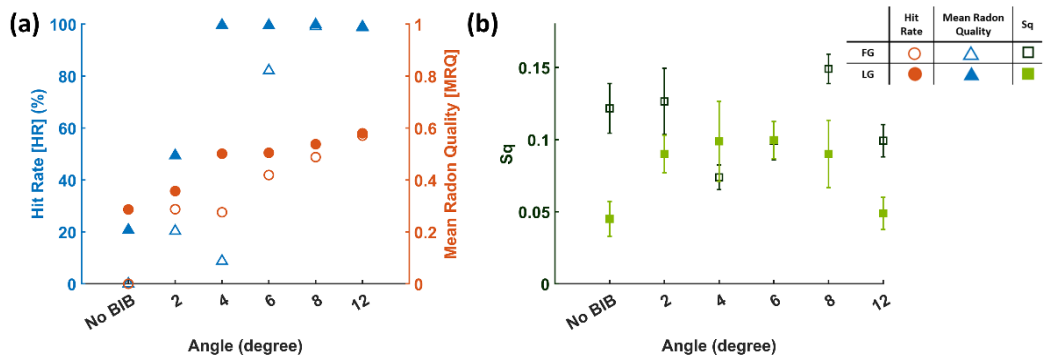

Figure 4 Quantitative analysis on the effect of gun tilt angle for broad ion beam polishing quality. (a) EBSD Hit rate and mean radon quality; (b) Surface Roughness (Sq) of samples surface after broad ion beam polishing with different angles of large grain (OPS polished) and fine grain (ground to 2000 grit using SiC) Zircaloy-4. Recipe: 8keV, no modulation, 1 rpm, 15 mins, 2-12° beam angle, room temperature. (In figure 2(b), the LG 2° is polished for 30mins).

2. **Voltage trials**

To understand the influence of voltage, tests on fine and large grain samples were carried out, varying the voltage from 2keV to 8keV, as detailed in Table 1 and Figure 3 shows the representative FSD and EBSD maps for large grains and only FSD and maps for fine grains. The hit rate and radon quality in Figure 4 show that the fine grain samples cannot be indexed between 2keV and 6keV.

For large grain samples, a higher angle results in a higher indexing success, where a gun voltage of ≥6 keV is sufficient to achieve >99% indexing success. For fine-grain samples, indexing is only possible at 8 keV, with a hit rate of 82.1%.

The FSD images show the difference in the texture caused by different voltages and grinding scratches are visible within the fine grain samples. Although the distribution of pits is still linear on the 8 keV polished fine grain sample, the surface roughness/texture is completely different from that under lower voltage. For large grains, the pits are denser and deeper with increasing polishing voltage.

In the EBSD maps of large grain samples, poor indexing is found within a grain, as seen in the 4keV polished sample. In the higher voltage maps, the obvious difference in the hit rate among different grains can be found. Points that cannot be indexed still correspond to secondary phase particles, which can be clearly seen in FSE images of 6 keV and 8 keV.

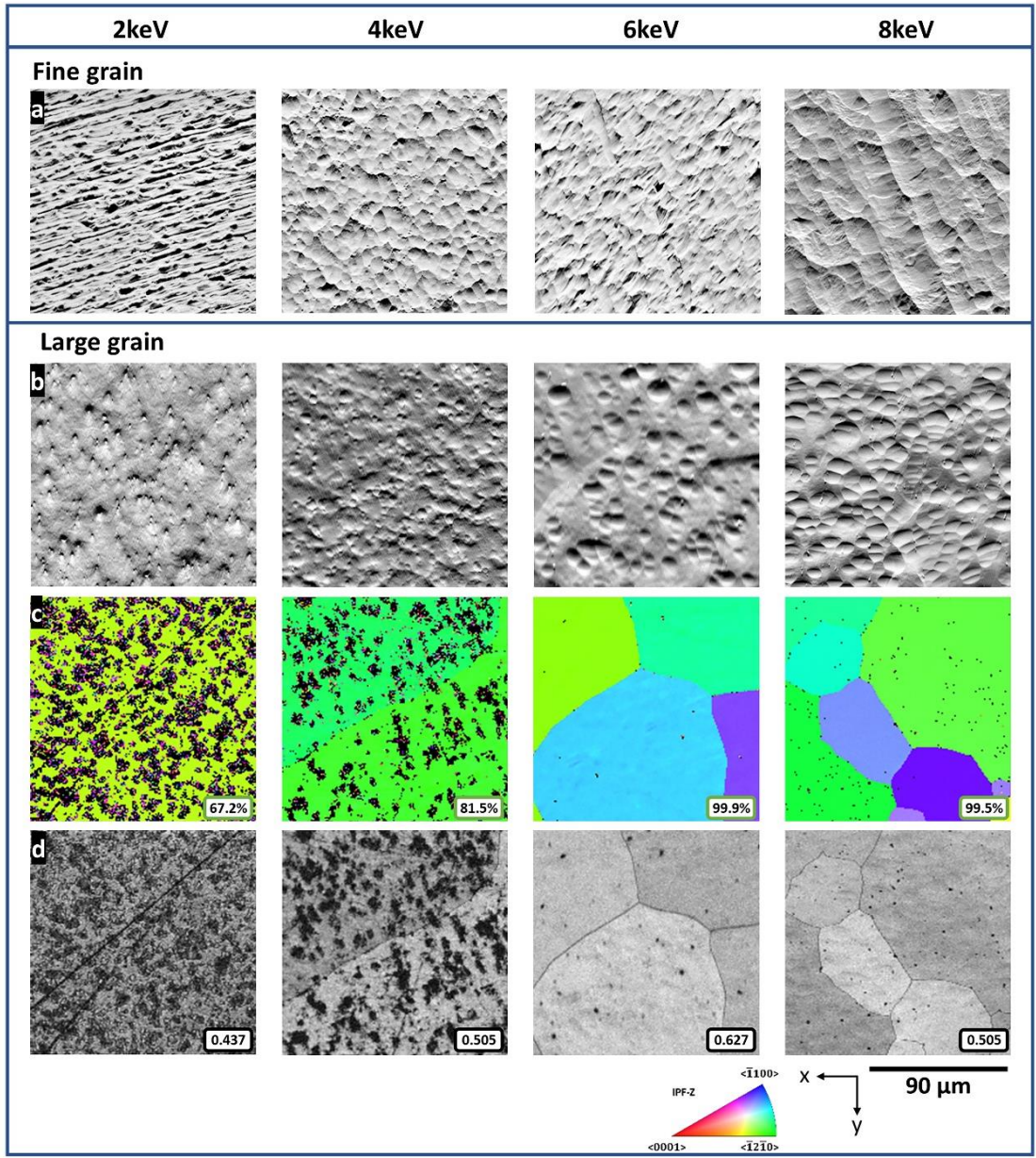

Figure 5 Effect of voltage on broad ion beam polishing quality for forescatter imaging of large grain (OPS polished) and fine grain (ground to 2000 grit using SiC) Zircaloy-4, and the successful EBSD mapping and radon quality for LG samples. The numbers in the green box are the percentage of successfully indexed points, and the numbers in the black box are the mean radon quality. Recipe: 6 °, no modulation, 1 rpm, 15 mins, 2-8 keV, room temperature. (a)(b) Forescatter imaging, (c) successful EBSD mapping

*with IPF-Z colouring and non-indexed points in black, (d) normalised pattern quality.*

*[The magnification is the same for all the micrographs in this figure.]*

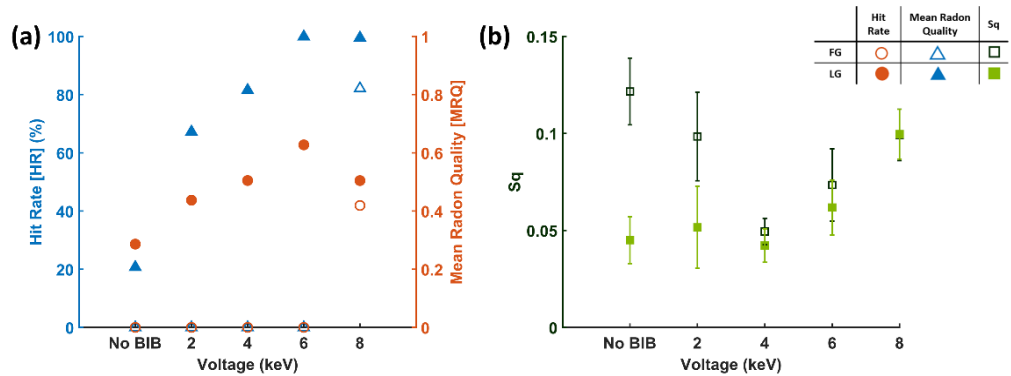

*Figure 6 Effect of voltage on broad ion beam polishing quality (a) EBSD Hit rate and mean radon quality; (b) Surface Roughness (Sq) of samples surface after broad ion beam polishing with different voltages of large grain (OPS polished) and fine grain (ground to 2000 grit using SiC) Zircaloy-4. Recipe: 6°, no modulation, 1 rpm, 15 mins, 2-8 keV, room temperature.*

### 3. Duration trials

Based upon the angle and voltage trials, it is easier to polish the large grain samples. This motivated exploration of how to improve polishing for the fine grain samples. The total polishing duration was extended from 3 minutes and 45 seconds (225 s) and then imaging and mapping were performed. The sample was then repolished using 2x the time of the previous polishing step, resulting in the final polishing having a total exposure to the BIB of 60 minutes.

This enabled sampling of the surface quality 3 min 45 s, 7 min 30 s, 15 mins, 30 mins and 60 mins and representative FSD and EBSD maps are shown in Figure 3, and the hit rate, radon quality, is marked in the corner of each figure.

The FSD images show that the basic shape of the dimple textures does not change apparently, but the size and area increase. The edge of dimples and 'ridgeline' is flatter with a longer polishing time within 60 minutes polishing.

Overall, the average EBSD pattern quality and radon quality increases with increasing polishing duration, and after 7min30s, the hit rates achieve >99% indexing.

The shape and size of grains in the EBSD maps are still similar between 3min45s and 7min30s, although the quality of the 3min45s polishing is poor with many misindexed points. While the indexing is highly successful after 7 min 30 s (>99%), the pattern quality maps indicate that the grain structure is clearer when the material is polished longer, and more material is removed. Material removal is indicated by comparing the microstructure between polishing steps, as new grains emerge in the map.

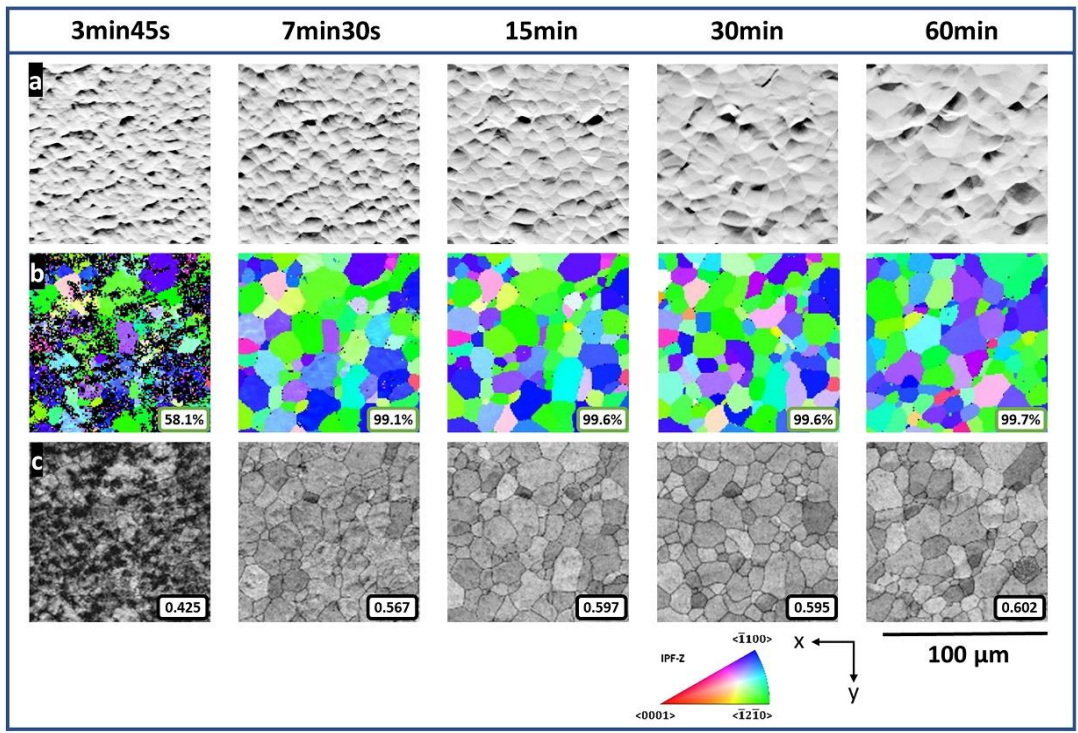

*Figure 7 Effect of total polishing duration on broad ion beam polishing quality for forescatter imaging and the successful EBSD mapping of fine grain (ground to 2000 grit using SiC) Zircaloy-4. The numbers in the green boxes are the percentage of successfully indexed points, and the numbers in the black boxes are the mean radon quality. Recipe: 8 °, 8 keV, no modulation,1 rpm, 3 min 45 s ~ 60 mins, room temperature. (a) forescatter imaging, (b) successful EBSD mapping with IPF-Z colouring and non-indexed points in black, (c) normalised pattern quality. [The magnification is the same for all the micrographs in this figure.]*

## 4. Cold PECS

To investigate the effect of polishing while the sample is cooled the system was cooled using liquid nitrogen, with the system temperature approximately -150 °C.

Due to the measurement location and differing cooling rates, the exact sample temperature is unknown but likely to be between -50 and -150°C.

Two comparison experiments were carried out, firstly to look at whether polishing cold affects the polishing quality (on fine and large grain samples), and secondly to expand this to look at the effect of voltage on fine grain samples (in comparison with room temperature). In both cases, results are compared with the equivalent sample and polishing conditions at room temperature.

Representative FSD and EBSD maps from the initial cold polishing trial are shown in Figure 8. For both grain sizes, cold polishing produces a well-indexed EBSD map with higher MRQ than the room temperature polish. There is also a visual difference in the FSD images for large grain, with a reduction in grain boundary shadowing and some highlighting of the SPPs. Measurement of the surface roughness (not shown) also showed a significant difference, with a change in Sq from 0.15±0.01 (RT) to 0.10±0.01 (cold) for fine grain and a reduction from 0.09±0.02 (RT) to 0.05±0.01 (cold) for large grain.

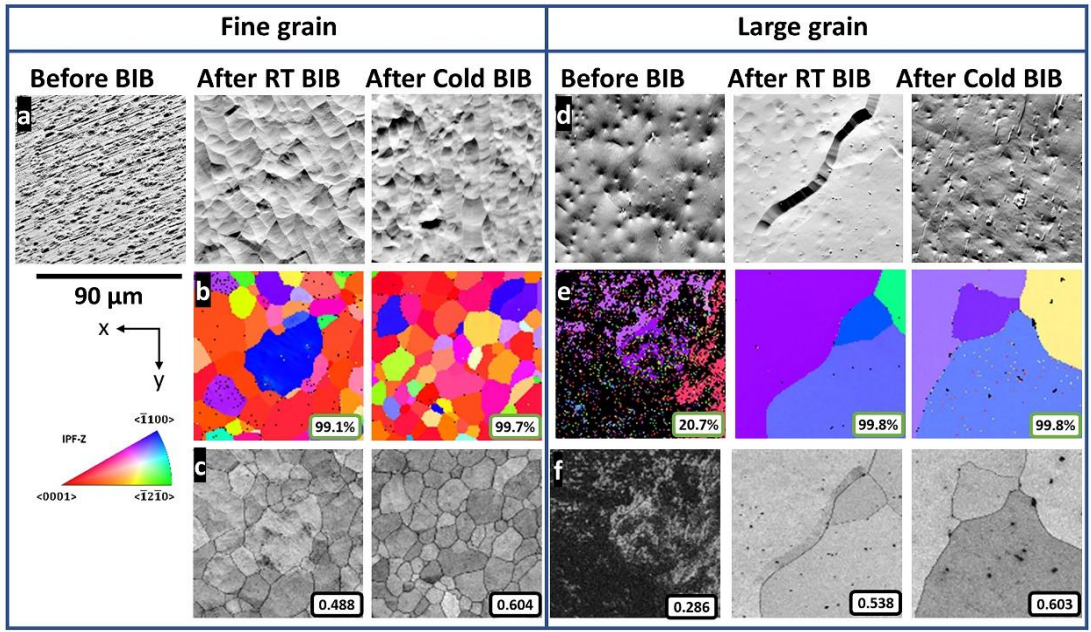

Figure 8 Effect of temperature on broad ion beam polishing quality for successful EBSD mapping and forescatter imaging of large grain (OPS polished) and fine grain (ground to 2000 grit using SiC) Zircaloy-4. The numbers in the green boxes are the percentage of successfully indexed points, and the numbers in the black boxes are the mean radon quality. Recipe: 8°, 8 keV, no modulation, 1 rpm, 15 mins, room temperature and cold broad ion beam polishing. *[The magnification is the same for all the micrographs in this figure.]*

Representative FSD and EBSD maps from the cold voltage trial are shown in Figure 9 (for comparison with room temperature, see Figure 5). Assessment of the hit rate, radon quality, and the surface roughness is presented in Figure 10 (see Figure 6 for the comparison room temperature results).

All cold voltage samples had high indexing success, with hit rates >96% at voltages ≥4 keV. Hit rate, MRQ and surface roughness (Sq) are all relatively stable at ≥4 keV, with no appreciable improvement with increasing voltage.

There is a significant difference between room temperature and cold for the voltage trials. Where the room temperature was only indexable at 8 keV (6°), at 82.1% indexed, higher indexing values are possible at all voltages under the cold conditions, with the hit rate varying from 39.6% to 98.9%. The FSD shows varying surface quality in the cold polished samples, with clear scratches visible in both the 2 keV and 4 keV samples. At higher voltages, the texture appears more uniform with no visible scratches. The 6 keV surface appears to be 'pitted' whereas the 8 keV surface closely resembles that seen in successful polishing at room temperature (see e.g., Figure 7).

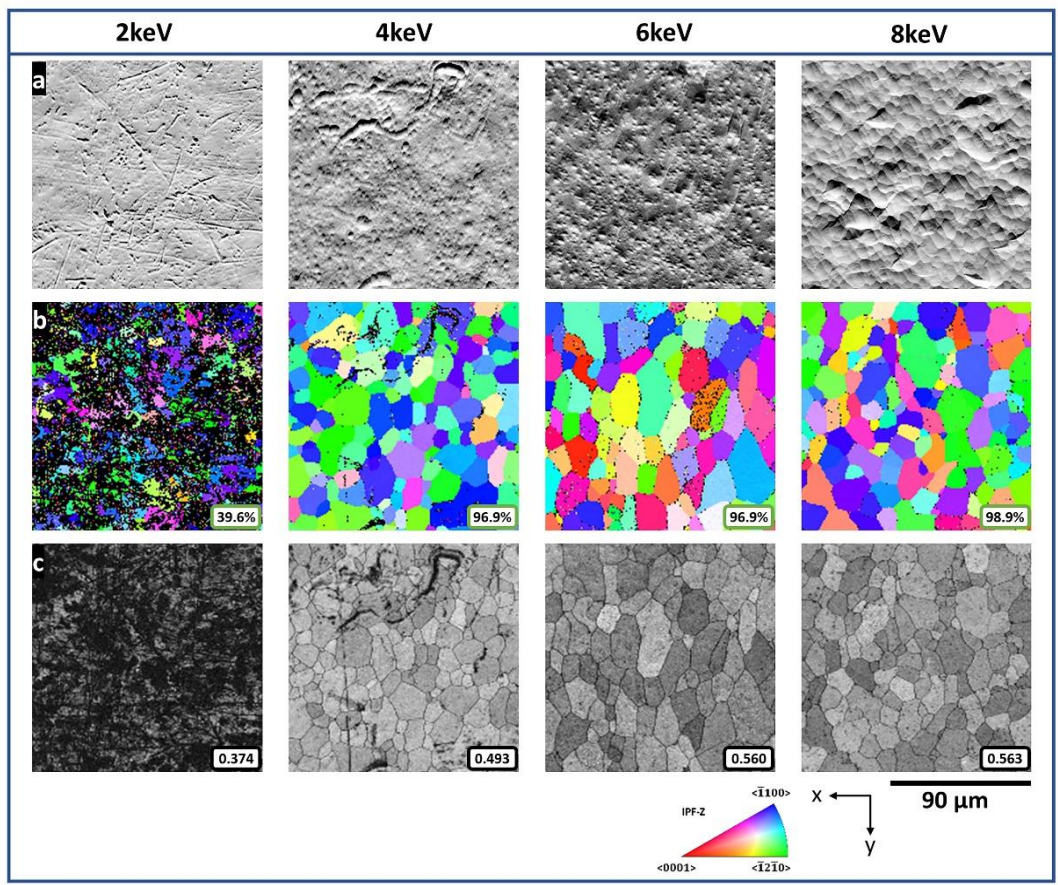

*Figure 9 Effect of voltage on broad ion beam polishing quality carried out at low temperature using fine grain (ground to 2000 grit using SiC) Zircaloy-4 samples. The numbers in the green boxes are the percentage of successfully indexed points, and*

*the numbers in the black boxes are the mean radon quality. Recipe: 6°, no modulation, 1 rpm, 15 mins, 2-8 keV, cold polishing. (a) Forescatter imaging, (b) successful EBSD mapping with IPF-Z colouring and non-indexed points in black, (c) normalised pattern quality. [The magnification is the for all the micrographs in this figure.]*

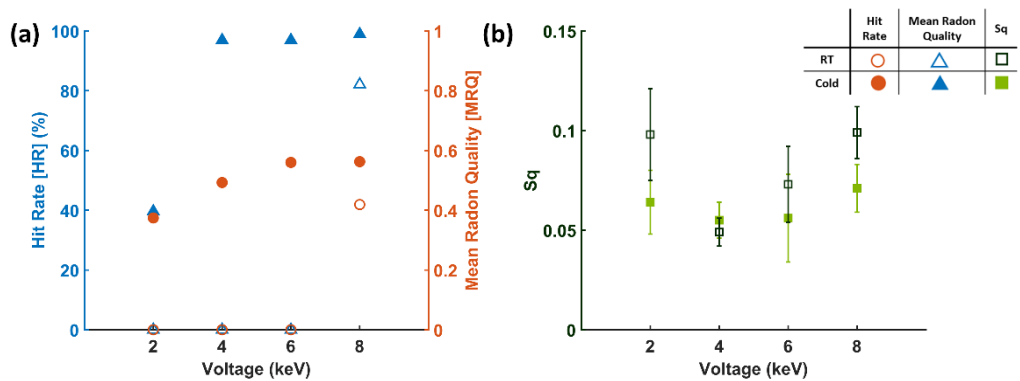

*Figure 10 Effect of voltage on broad ion beam polishing quality at room temperature and cold (a) EBSD Hit rate and mean radon quality; (b) Surface Roughness (Sq) of samples surface after broad ion beam polishing with different voltages and temperature of fine grain.*

**5. Grain orientation**

There exist various practical aspects that decide the topographical quality of the sample on parameters, and grain orientation is a vital factor.

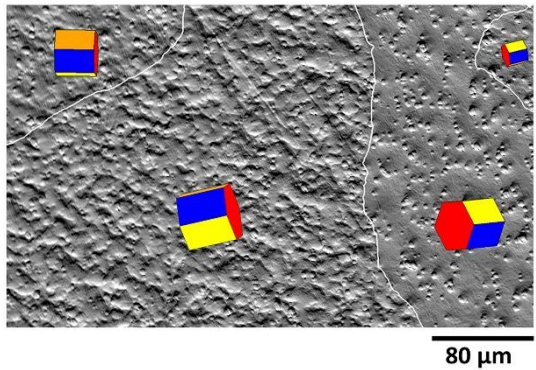

*Figure 11. Anisotropic polishing based upon crystal orientation of the matrix. Recipe: 6°, no modulation, 1 rpm, 15 mins, 4 keV, room temperature.*

A variation in polishing with grain orientation has been observed in Figure 11. The surface texture and roughness vary with different grain orientations. The grains with the <c> axis out of the surface of the sample were polished more easily and indexed was also found in the 2° fine grain trial in Figure 3. The rest of the sample is unindexed or poor indexed. Furthermore, this variation in polishing with grain orientation will give rise to grain-to-grain changes in topography depending on the misorientation between two neighbouring grains.

## IV. Discussion

These results show that BIB can be used to directly produce high quality surfaces suitable for EBSD analysis from ground samples, significantly increasing throughput for sample preparation. This was observed with the fine grain samples, which are polished until 2000 grit silicon paper. The large grain samples are after OPS polishing, which were easier to index and further aimed to find out the surface texture and conditions. The analysis from large grain samples also highlighted the differences in cold polishing when there are requirements on grain boundaries and removal of SPPs.

Among all these trials, 8° and 8keV are considered to be the best polishing angle and voltage. Cold polishing is suggested to do if it can be done, except if the grain-to-grain relief is required for the following analysis.

From the angle trials, it can be found that a higher than 8° leads to a high spatial frequency increase in surface roughness; while an angle lower than 6° results in not enough grains polishing with significant grain surface normal effects. However, it should be noticed that different angles cause the changes in the shape of the probe, which will change the area of polishing – with a shallow angle distributing the polishing over a larger area (likely reducing dose).

Lower voltage reduces current, reduces dose, and reduces the sputtering and ion penetration depth, while the control in a BIB is not independent. This was accentuated when we tried to produce surfaces suitable for EBSD analysis from the as ground-polished condition, where 8 keV is required to achieve this. Note that 8 keV is set as the highest voltage limitation on the instrument among all these trials.

Different grain sizes and the presence of SPP will also change the balance of BIB parameters. This will create systematic misindexing or preferentially induce shadowing for specific features.

Parameter optimisation of sample preparation can easily include multiple independent parameters (grain size, final mechanical polishing conditions, angle, voltage, temperature, time) and our results show that some of these are coupled. This makes it easy to consider a range of experiments, but the number of samples we would have to create is prohibitive. For example, a group of tests on fine grain sample that was mechanically polished with OPS can also be done, but we did not have the time or scope to do this, and the ground polished fine grain samples already index well.

In our work, we did not explore the use of modulation (which involves turning the guns on and off during rotation) and the speed of the rotation. For a single phase metallic sample, this is unlikely to make a significant impact, but this could be useful for other samples. One example suggested by the manufacturer, is using zero rotation for an aligned single crystal, e.g. where there are specific crystallographic effects for milling, or samples with directional surface roughness where shadowing can be accentuated. Another example where modulation could be helpful is during cross sectioning, where the rotation speed is synchronised with the gun power to reduce polishing on the back face of the shrouded material.

In this study, Evaluation of the gun set up, focus, and cleanliness, as well as the impact of accelerating voltage on ion current hints that although scientifically we might consider varying parameters to enable us to control total dose, the equipment as supplied is not configured to do this.

For a physics informed understanding of ion-matter interactions during BIB polishing, it is helpful to consider: the sputter yield, isotropy of sputtering across the polycrystalline and multi-phase sample, as well as an understanding of sample preparation induced defect generation. Adjusting the voltage, gun tilt angle, polishing duration, and polishing temperature are necessarily indirect methods of affecting these processes, and therefore influence the quality of our prepared surface.

The sputter yield is the mean number of atoms removed from the base material per incident ion[29], and in general the yield increases with a larger angle of incident ions (0°~18°)[30]. The sputtered depth also has a linear relationship with the bombardment time (i.e. increases with total dose). The sputter depth can be increased if the dose per unit area increases, e.g. with a higher current density per unit area of the incident ions[31].

The effect of varying incident ion voltage on sputter yield is more complicated. Simon outlines three regimes: a threshold at low energy (<50 eV) with limited sputtering; knock-on for ~10 eV to 1 keV where the surface can be dislodged and some atoms can be ejected resulting in a linear dependence of voltage on yield and sputter yields of 1-3 (ejected atoms per incident ion)[32]. When the ion voltage is above 1 keV, there is a collision cascade where each incident ion can dislodge more than one atom and generate multiple point defects in the target as well as potentially resulting in multiple sputtered atoms/ions being released, and this results in a ~flat sputter yield with increasing voltage[32]. The precise thresholds for these regimes will depend on the bombarding ion species, the surface morphology, and the target crystal structure and orientation.

The combination of these effects, especially in the collision cascade region, makes it extremely difficult to predict *a priori* sputter rate and defect generation rate. This is complicated further by ion channelling effects (which result in anisotropic behaviours based upon crystal orientation and phase, see Figure 11). Furthermore, the role of sample temperature is likely important in controlling the ratio of (permanent) defect generation and sputtering, and here we find that sputtering while cold at high voltage is especially useful. The surface morphology can play a further effect, due to redeposition of the sputtered material affecting the removal rate and the final surface quality.

For practical BIB polishing, we also note that the stability of the instrument is also important. For example, we have found that may be fluctuations in the ion current from the guns.

Together these factors mean that in this manuscript, we have opted to focus on the more practical aspects of BIB polishing and systematically explore variables we can access with the instrument. We have been informed in the selection and sampling of accessible parameter space through the physics-informed understanding of each of mechanisms of sputtering and defect generation.

**Recommendations**

If you would like to use BIB to polish your specimens, we suggest that first you find a sample that is close in grain structure and chemistry to an existing recipe. Next, we suggest adjusting the voltage such that the surface is polished (as too low a voltage results in no material removal) and even polished for too long a duration, i.e. resulting in etching of the surface. Then you can reduce the voltage until the indexing threshold is achieved. Once you know this voltage, you can optimise the angle of polishing to optimise surface roughness. We caution that this can be counterintuitive, that smoother samples with less high-frequency BIB surface roughness (i.e. ~uncorrelated with the microstructure) may not index as well as those with this high-frequency roughness. Once the angle is optimised, you can re-check the threshold voltage, and finally optimise for the polishing duration. We caution that the polishing angle and voltage can be coupled, i.e. a sample polished with a lower angle and a lower voltage can be polished as well as a sample with a lower angle and lower voltage, as observed for our large grain samples.

It is worth noting that these parameters are linked to the microstructure in terms of relative sputtering rate of the different features, and so may need to be changed slightly if the microstructure is changed, e.g. cold work, change in precipitate density, or changes in chemistry.

In zircaloy-4, we observe that the best parameters used at room temperature can be used when the sample is chilled with LN2, and the indexing is improved. It is easier to vary parameters without also having to chill the sample for each trial.

As a final item of advice, we strongly suggest that each user learns to well align and clean the ion guns prior to experimenting with their polishing routine.

## V. Summary

This project provides practical guidance and exploration of parameters for BIB polishing with PECS. For zircaloy-4, it proves that the surface of ground polished samples can reach a satisfying state after BIB polishing, without further mechanical and chemical polishing. The surface indexing rate increases with both gun tilt angle and beam voltage. The cold BIB polishing plays an important role in keeping a flat surface and decreasing the roughness. When performed at a cold temperature, the BIB method is more complicated and time-consuming but can show a better surface quality, especially in removing the artefacts.

**Acknowledgment**

We would like to thank Dr Mahmoud Ardakani for his assistance in fixing the PECS instrument during our experiments.

**Author Contribution Statement**

NF and RB prepared the samples, devised the PECS trials and ran the instrument, as well as performing the EBSD and SEM experiments. NF performed the surface roughness measurements. NF led the drafting of the manuscript, with input from all authors during the writing. All authors contributed to the final manuscript. We note that if required, either of the two first authors may cite this work with either of their names as the first name listed.